\documentclass[aps,twocolumn,showpacs,superscriptaddress]{revtex4}
\usepackage{graphicx}
\usepackage{subfigure}
\usepackage[latin1]{inputenc}
\usepackage{amsmath}
\usepackage{amssymb}
\usepackage{color}
\usepackage{comment}

\newcommand{\nc}{\newcommand}
\nc{\bra}[1]{\langle #1|}
\nc{\ket}[1]{|#1\rangle}
\nc{\braket}[1]{\left\langle #1 \right\rangle}
\nc{\equ}[1]{\begin{eqnarray*}#1\end{eqnarray*}}
\nc{\equn}[1]{\begin{eqnarray}#1\end{eqnarray}}
\nc{\dagg}{^{\dagger}}
\nc{\conj}{^{*}}
\nc{\dx}[1]{\, \mathrm{d} {#1}}
\nc{\Dx}[1]{\mathcal{D} {#1} \,}
\nc{\la}{\langle}
\nc{\ra}{\rangle}
\nc{\Tr}{\text{Tr} \,}
\nc{\e}{\text{e}}
\nc{\Id}{\mathbb{1}}
\nc{\eps}{\varepsilon}
\nc{\der}[2]{\delta \frac{{#1}}{\delta {#2}}}
\nc{\pder}[2]{\frac{\partial {#1}}{\partial {#2}}}
\nc{\bigO}{\mathcal{O}}
\nc{\eq}[1]{Eq.(\ref{#1})}
\nc{\chap}[1]{Chapter \ref{#1}}
\nc{\sect}[1]{Section \ref{#1}}
\nc{\fig}[1]{Fig.\ref{#1}}
\nc{\Fig}[1]{Figure \ref{#1}}
\nc{\tabl}[1]{Table \ref{#1}}
\nc{\app}[1]{Appendix \ref{#1}}

\nc{\eg}{\emph{e.g.} }
\nc{\ie}{\emph{i.e.} }

\nc{\alg}[1]{\textcolor{blue}{#1}}

\begin{document}

\noindent {\bf Comment on:  ``Observation of Locally Negative Velocity of Electromagnetic Field in Free Space''}

\vspace{5mm}
In a  very interesting Letter \cite{Budko_2009_1}, and in related publications \cite{Budko_2009_2,Budko_2011},  the near- and intermediate- as well as far-field causal properties of classical electro-magnetic fields have been discussed in great detail by making use of  a representation of a Greens function where  the  various spatial dependencies of the components of the electric field are explicit, i.e., 
%
%
\begin{align}  
%
%
\nonumber 
    E_k({\bf x},t) = -\frac{1}{4\pi \epsilon_0} \int_D d^3x'  
(\partial_k \partial_n'\frac{1}{|{\bf x}-{\bf x}'|})\int_{t_0}^{t_R}dt'J_n({\bf x}',t')\\
\nonumber
- \frac{1}{4\pi \epsilon_0c}\int_D d^3x' |{\bf x}-{\bf x}'|(\partial_k \partial_n'\frac{1}{|{\bf x}-{\bf x}'|})J_n({\bf x}',t_R)\\
+ \frac{1}{4\pi \epsilon_0c^2}\int_D d^3x'\frac{1}{|{\bf x}-{\bf x}'|}(\theta_k\theta_n -\delta_{kn})\partial_tJ_n({\bf x}',t_R)\,\, .
\label{eq:budko_1}
%
%
\end{align}
%
%
Here  $t_R= t- |{\bf x}-{\bf x}'|/c$  is a retarded time-variable and $t_0$ a suitably chosen initial time such that electric charge density $\rho({\bf x}',t')$ vanish for $t'=t_0$.  Furthermore,  $\theta_n = (x_n- x_n')/|{\bf x}-{\bf x}'|$ is a component of a unit vector  in terms of   the observation point ${\bf x}$  and a source point ${\bf x}' \in D$ in the domain $D$ of the current source $J_n({\bf x}',t')$. The partial derivatives $\partial_k$  and $\partial_n'$ are acting on the ${\bf x}$ or  the ${\bf x}'$ dependence, respectively. By performing the partial derivatives $\partial_k \partial_n'$  in Eq.(\ref{eq:budko_1}) it  is seen  that Eq.(\ref{eq:budko_1}) has exactly  the same form as used in Ref.\cite{Budko_2009_1}. The non-local time dependence in the first term of Eq.(\ref{eq:budko_1}) is due to the elimination of a dependence of the charge density $\rho({\bf x}',t')$ in the electric charge conservation law $\partial_t\rho({\bf x}',t') + \partial_kJ_k({\bf x}',t') = 0$. It was noticed  in Ref.\cite{Budko_2009_1}   that the representation Eq.(\ref{eq:budko_1}) can lead to locally negative velocities and apparent superluminal features of electro-magnetic fields  also demonstrated  experimentally \cite{Budko_2009_1,Budko_2011}. These observations do not challenge our understanding of causality since they describe phenomena that occur behind the light front of electro-magnetic signals (see, e.g., Refs.\cite{Lautrup_2001,Budko_2011} and references cited therein). It was, however,  also remarked in Ref.\cite{Budko_2009_1} that the derivation of Eq.(\ref{eq:budko_1})  "{\sl appears if potentials are linked by the relativistic Lorentz gauge, which is something to think about}''.   We find the content of this quotation misleading. The observable electric field as in Eq.(\ref{eq:budko_1})  should, of course, be gauge invariant.    By inspection of the actual mathematical derivation as referred to (in particular Chapter 26 of  Ref.\cite{Hoop_1995}), one indeed observes  that the Lorentz gauge is used.   Now Eq.(\ref{eq:budko_1}) appears to be different from the standard and gauge-invariant expression in terms of retarded  charge- and current-densities (see, e.g., Chapter 6.5 in  Ref.\cite{Jackson_1999}), i.e., 
\begin{align}  
%
%
\nonumber
     E_k({\bf x},t) = - \frac{\partial}{\partial t} \bigg ( \frac{1}{4\pi \epsilon_0 c^2 } \int d^3x' \frac{ J_k({\bf x}',t_R) }{|{\bf x}-{\bf x}'|} \bigg ) \\
     -  \frac{1}{4\pi \epsilon_0} \int d^3x' \frac{\partial_k'\rho \big ( {\bf x}',t_R \big )}{|{\bf x}-{\bf x}'|} ~ ,
\label{eq:ave_electric}
%
%
\end{align}
where $\partial_k'\rho ({\bf x}',t')$ in Eq.(\ref{eq:ave_electric}) has to be evaluated for a fixed value of $t'= t_R$ . 
By making use of current conservation and elementary partial integrations one finds, after some remarkable and non-trivial cancellation of terms, that Eqs.(\ref{eq:budko_1}) and (\ref{eq:ave_electric}) are,   actually equivalent apart from possible boundary terms that vanishes if, e.g.,  $J_n({\bf x}',t')$ goes to zero sufficiently fast at the boundary of $D$.  Furthermore, the gauge-independence of Eq.(\ref{eq:ave_electric}) has  been discussed in various contexts (see, e.g., Ref.\cite{Jackson_2002} and references cited in Ref.\cite{Skagerstam_2019}). With regard to another issue as raised by Budko \cite{Budko_2009_1} on the role of Eq.(\ref{eq:budko_1}) in quantum optics,  we notice at least the following fact.  Interpreted as an expectation value of second-quantized physical degrees of the electro-magnetic  field  in the presence of arbitrary but classical charge- and current-densities sources, Eq.(\ref{eq:ave_electric})
is obtained by exactly  solving for the corresponding quantum-mechanical equations of motion (see Ref.\cite{Skagerstam_2019} and references cited therein). As long as characteristic dimensions of the sources are small compared to any typical wave-length of photons, this should be an exact result according to the analysis of the infrared problem in quantum field theory. In view of these remarks, we do not find that the near- and intermediate-field terms as exhibited explicitly in Eq.(\ref{eq:budko_1}) are  by "far less understood in quantum optics" but, nevertheless they deserves more attention in a quantum-mechanical context. 

\vspace{2mm}
\noindent B-S. K. Skagerstam\\
Department of Physics\\
NTNU\\
N-7491 Trondheim\\
Norway\\
PACS numbers: 03.50.De, 41.20.Jb


\end{document}